\documentclass[twocolumn,,amssymb,amsmath,aps,showpacs,floatfix,nofootinbib,showpacs,balancelastpage]{revtex4}
\usepackage{amsmath}
\usepackage{amssymb}

\usepackage{hangcaption,fancybox,feynmp}
\usepackage{indentfirst}
\usepackage{graphicx}
\usepackage{epsfig,subfigure,psfrag}
\usepackage{mathrsfs}

\begin{document}

\title{\boldmath Distinguishing the  Color Octet Axial-Vector-like Particle for Top Quark
 Asymmetry via Color Flow Method at the LHC }

\author{
Tianjun Li$^{1,2}$ %\footnote{E-mail:tli@itp.ac.cn},
Xia Wan$^{3}$ %\footnote{E-mail:xia.wan@pku.edu.cn},
You-kai Wang$^{1}$ \footnote{E-mail:wangyk@itp.ac.cn},
and Shou-hua Zhu$^{3,4}$ %\footnote{E-mail:shzhu@pku.edu.cn}
}

\affiliation{
$ ^1$ State Key Laboratory of Theoretical Physics
and Kavli Institute for Theoretical Physics China (KITPC),
Institute of Theoretical Physics, Chinese Academy of Sciences,
Beijing 100190, P. R. China \\
$ ^2$School of Physical Electronics, University of Electronic Science
and Technology of China, Chengdu 610054, P. R. China\\
$ ^3$ Institute of Theoretical Physics $\&$ State Key Laboratory of
Nuclear Physics and Technology, Peking University, Beijing 100871,
China \\
$ ^4$ Center for High Energy Physics, Peking University,
Beijing 100871, China }

\date{\today}

\begin{abstract}

The excess in top quark forward-backward asymmetry has been a hot topic in recent years.
Although there are many proposals to explain it,  most of them can not fit the differential
distributions well. The color octet axial-vector like particle, with mass near
the top quark pair threshold, is still a good description of the differential distributions.
We study how to distinguish the color singlet and octet mediating particles in top quark
pair production by adopting the color flow method. For the first time, we show that
such a proposal for the froward-backward asymmetry can be cross-checked indirectly at the LHC.

\end{abstract}


\maketitle

\flushbottom

%\section{Introduction}
%\label{Introduction}
{\bf Introduction:} Being the heaviest particle in the Standard Model (SM), the top quark is
considered to play a special role in the electroweak (EWK) symmetry breaking. Precise
understanding of the top quark production processes is essential for the study of physics
beyond the SM since they are usually the most important background of these new physics.
Due to the short lifetime of top quark, its spin is fully inherited by its decay products.
Thus, top quark can be used as a tool to study the spin information of the related processes.
Similarly, we study in this paper that the top quark can also be utilized to study
the color properties of the production processes. An efficient treatment is just to
analyze the shape of the b jet from the top decay, as the top quark color are fully
carried on by its decay product b quark.

Recently, the anomaly of top quark forward-backward asymmetry~($A_{FB}$) has been a hot topic.
It was observed by the CDF and D0 Collaborations at the Tevatron and the CDF latest result
still supports that there is an excess in top $A_{FB}$~\cite{Aaltonen:2012it}. Disappointingly,
the top $A_{FB}$~(or sometimes named as the charge asymmetry) is very difficult to be measured
precisely at the LHC due to the large symmetric gluon-gluon fusion component. Two of us
have introduced a so called color octet axial-vector-like boson to explain
this anomaly~\cite{Xiao:2010ph}. Such a new particle is also called the light
axigluon~\cite{Gross:2012bz}. Recent studies showed the color octet axial-vector-like boson
can have the virtue to give the correct differential distributions, such as
 $d\sigma/d M_{t\bar{t}}$, $d A_{FB}/ d M_{t\bar{t}}$, and $d A_{FB}/d Y$, where $\sigma$ is
the production cross section, $M_{t\bar{t}}$ is the top pair invariant mass, and $Y$ is the
rapidity of the top quark. However, the other explanations often have difficulties to give
such a good fit~\cite{Wang:2011hc}. A very important feature of the new boson is that it carries
octet color and then can not be seen directly due to the color confinement. Interestingly,
the color flow in quark pair production does have been studied in Ref.~\cite{Gallicchio:2010sw}.
The mediating particles carrying different colors (color singlet or color octet) can be
distinguished successfully. In this paper, we adopt the color flow method in the
top $A_{FB}$ analysis, especially in studying the color octet axial-vector-like boson explanation.

%The paper is organized as follows: Section~\ref{Zc} is a brief introduction of the color octet axial-vector like boson and the %best fitted parameters. Section~\ref{colorflow} is the discerption of the color flow methods and its application on top %$A_{FB}$. The last section is our conclusion.

%\section{The color octet axial-vector-like boson\label{Zc}}
{\bf The Color Octet Axial-Vector-Like Boson:} The color octet axial-vector-like boson
$Z_C$ is an effective description to explain the top $A_{FB}$ anomaly.
The effective lagrangian can be written as
\begin{equation}
\begin{array}{rl}
{\cal L} \propto &- i g^Q_A \bar{Q} \gamma_\mu \gamma_5 T_a Q {Z_C}^\mu_a\\
& - i g^q_A \bar{q} \gamma_\mu \gamma_5 T_a q {Z_C}^\mu_a~,\\
\end{array}
\end{equation}
in which $Q=t, b$ and $q=u, d, c, s$.
The squared invariant amplitude with color and spin summed is
\begin{equation}
 \begin{array}{l}
 \sum\limits_{\text{Color, Spin}}|{\cal M}|^2= \frac{C_A C_F}{2}\{4g_s^4 (1+c^2+4m^2)\\\\
  +\frac{8g_s^2 \hat{s}(\hat{s}-M_{Z_C}^2)}{(\hat{s}-M_{Z_C}^2)^2+M_{Z_C}^2\Gamma_{Z_C}^2}2g_A^q g_A^t c\\\\
  +\frac{4\hat{s}^2}{(\hat{s}-M_{Z_C}^2)^2+M_{Z_C}^2\Gamma_{Z_C}^2}
 (g_A^qg_A^t)^2(1+c^2-4m^2)\},
 \end{array}
\label{eq.simp_axigluon}
\end{equation}
where $C_A=3$~, $C_F=4/3$, $g_s$ is the strong coupling constant, $m=m_t/\sqrt{\hat s}$, ${\hat s}$ is the parton level interacting energy, $\beta=\sqrt{1-4m^2}$, and $c=\beta\cos\theta$. The decay width $\Gamma_{Z_C}$ of $Z_C$ is given by
\begin{equation}
\Gamma_{Z_C}=\sum\limits_{i}^{2m_i<M_{Z_C}}\frac{g_A^i}{4\pi}\frac{C_F}{8}
(1-4m_i^2/M_{Z_C}^2)^{\frac{3}{2}}M_{Z_C}
\label{ZCwidth}
\end{equation}
In Eq.~(\ref{eq.simp_axigluon}), the first term is the contribution from SM gluon mediation,
 the second term is the interference between $q\bar{q}\to g \to t\bar{t}$ and
 $q\bar{q}\to Z_C \to t\bar{t}$, and the last term is the self-conjugation of the $Z_C$
mediating process. Some characters can be described for the $Z_C$ explanation:
 (1) The additional asymmetry comes from the interference between the
gluon and $Z_C$ processes. This can only change the angular distribution of
the produced $t\bar{t}$ events and has no contribution to
 the total $t\bar{t}$ cross section; (2) The third term of $Z_C$ self-conjugation can
contribute to the $t\bar{t}$ total cross section. However, this contribution can be
suppressed by adopting proper couplings;  (3) The Breit-Wigner propagator of $Z_C$
shows that $Z_C$ can only impact on its near mass peak region. Actually,
our original idea was to introduce a near top pair production threshold
particle to replace the usual heavy ($>1$~TeV) axigluon.

\begin{figure}[htbp]
\begin{center}
\includegraphics[width=0.23\textwidth]
{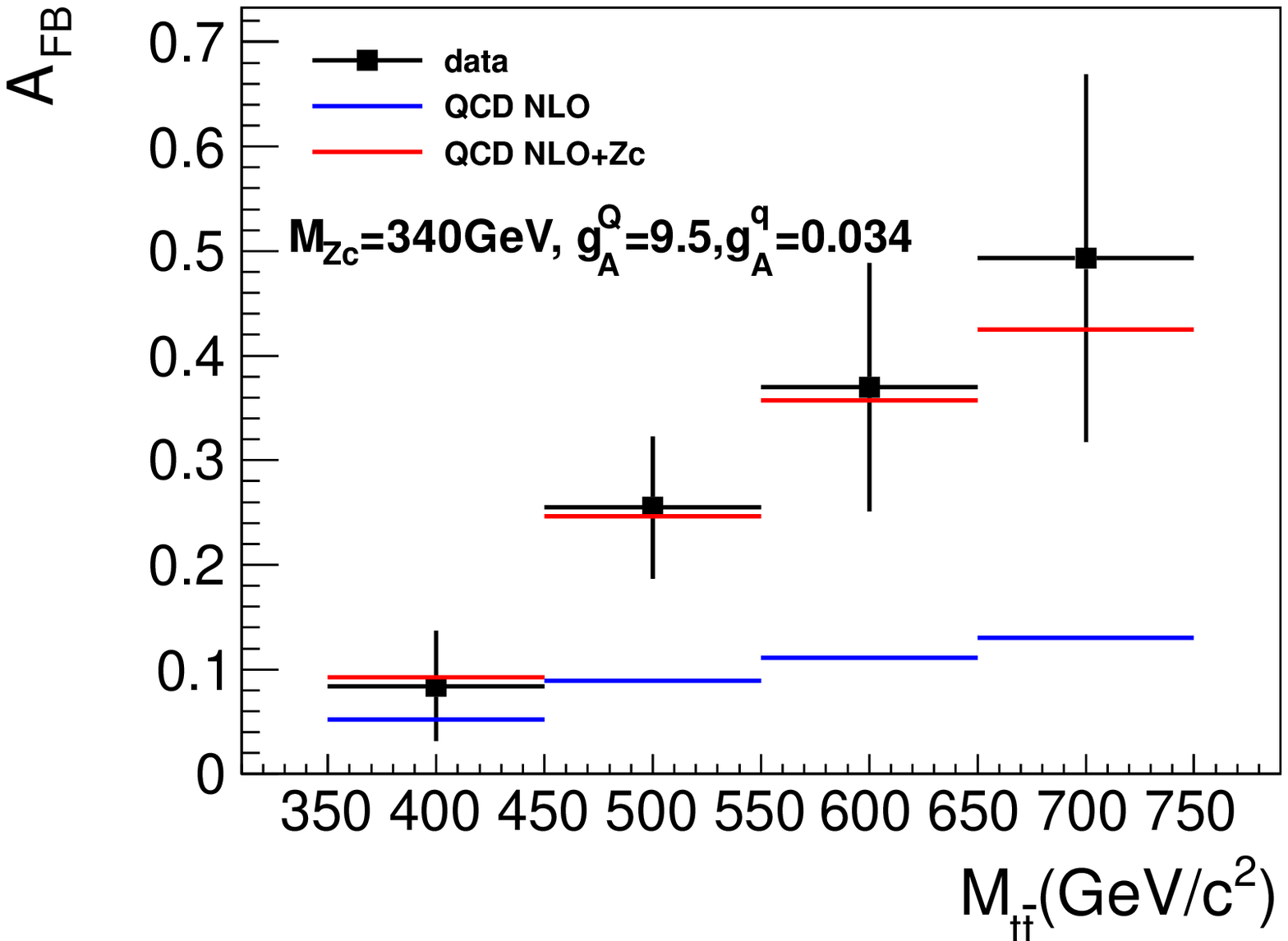}
\includegraphics[width=0.23\textwidth]
{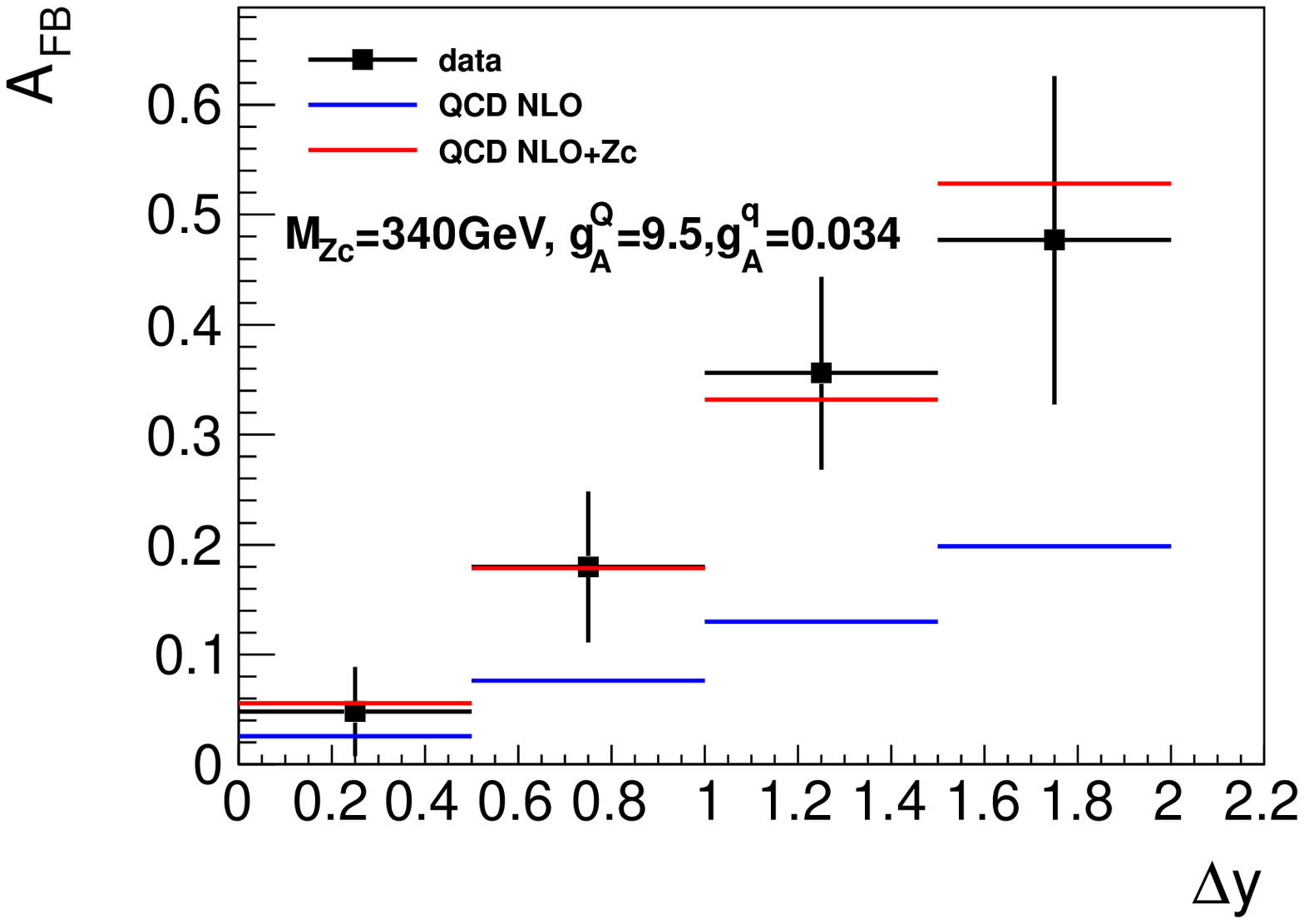}
\includegraphics[width=0.23\textwidth]
{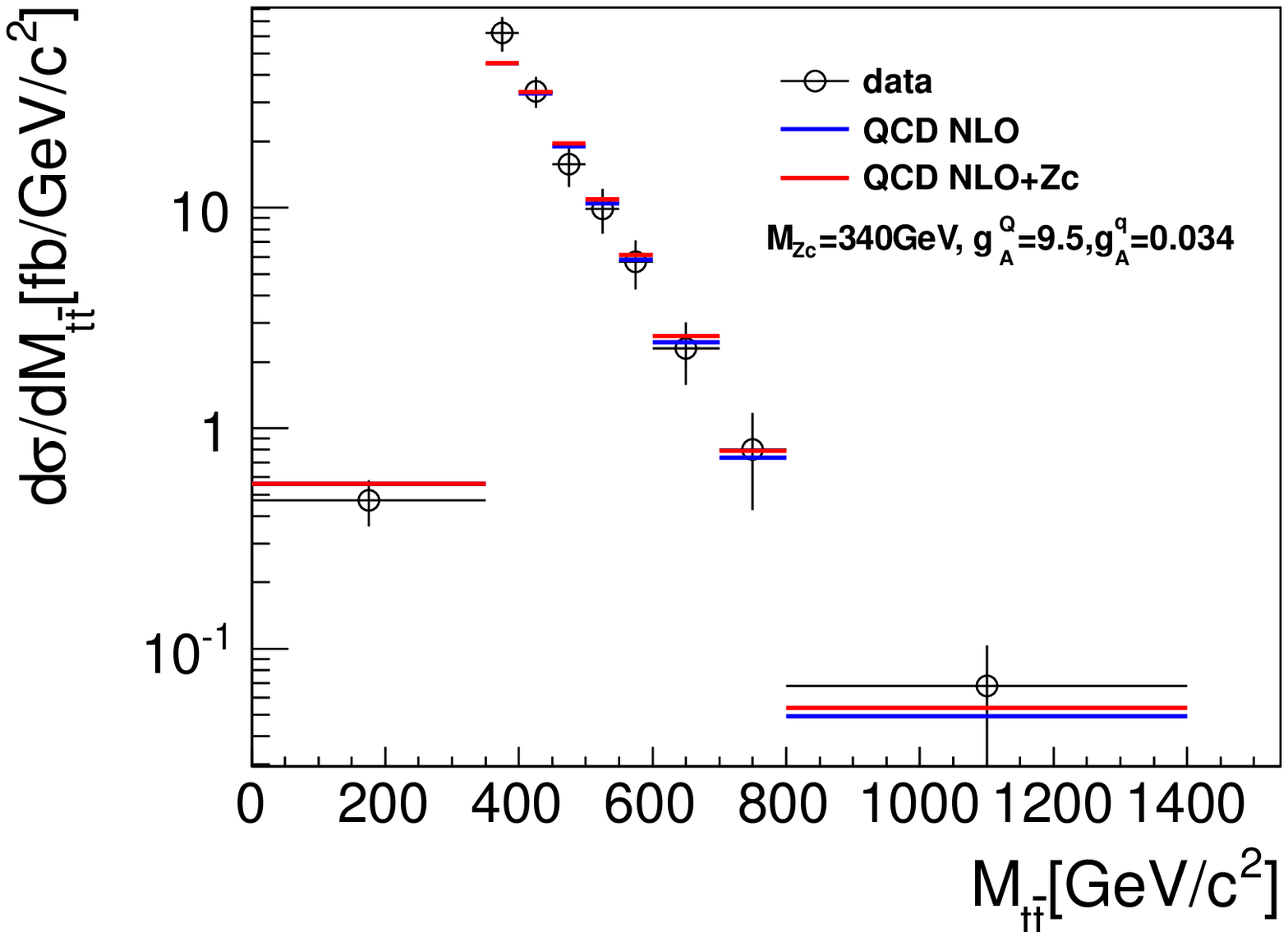}
\end{center}
\caption{\label{Zc-differential-distribution}The differential distributions with best fit $Z_C$ parameters. The experimental data comes from~\cite{Aaltonen:2012it, Aaltonen:2009iz}.}
\end{figure}

Fig.~\ref{Zc-differential-distribution} shows the best fit results by $Z_C$. The experimental data of $A_{FB}$ versus $M_{t\bar{t}}$ and $\Delta Y$, together with the differential distributions of the cross section $d\sigma/d M_{t\bar{t}}$ are constructed in the $\chi^2$ global fit. The free parameters to fit can be taken as $g_A^q$, $g_A^Q$ and $M_{Z_C}$, or equally as $g_A^q g_A^Q$, $\Gamma_{Z_C}$ and $M_{Z_C}$. Table~\ref{fitted-parameter} shows the consistent results in the two parametrization scenarios.

\begin{table}[htb]
\caption{\label{fitted-parameter}The best fit parameters in two free parameter selection scenarios. The unit of the mass and width is GeV. }
\begin{tabular}{ccc|ccc}
\hline\hline
$g_A^q$ & $g_A^t$ & $M_{Z_C}$ & $g_A^q g_A^t$&$\Gamma_{Z_C}$ &$M_{Z_C}$ \\
\hline
 0.034&9.5 & 340 &0.32 &400&340 \\
\hline\hline
\end{tabular}
\end{table}
It can be seen that $Z_C$ has tiny couplings to the light quarks and has very strong couplings to the heavy quarks. People may argue that $g_A^Q$ is so large that it may destroy the perturbation of the theory. Our statement is that the color octet axial-vector-like particle $Z_C$ is just an effective explanation of the top $A_{FB}$ anomaly. It is not necessary to be taken as a fundamental particle.

\begin{figure}[htbp]
\begin{center}
\includegraphics[width=0.3\textwidth]
{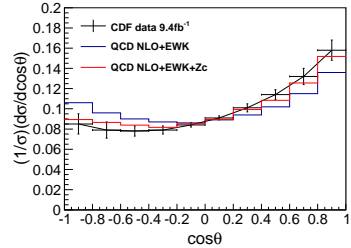}
\end{center}
\caption{\label{Zc-angular-distribution}The normalized angular distributions of the top quark in $t\bar{t}$ rest frame at the Tevatron. The experimental data and SM QCD NLO +EWK predicted values are taken from Ref.~\cite{angular-distribution}. }
\end{figure}

Actually, the forward-backward asymmetry is just a variate to exhibit the angular distribution of the final produced top quarks. For large enough luminosity, we can study the angular distribution directly and more information can be reserved by avoiding the half plane angular integration. Recently, by analyzing the $9.4~\mbox{fb}^{-1}$ data at the Tevatron, the CDF Collaboration gave the polar angular distribution of the top quark in the $t\bar{t}$ rest frame in Ref.~\cite{angular-distribution}.
Fig.~\ref{Zc-angular-distribution} shows our result by taking the above best fit parameters. The introduction of an effective color octet axial-vector-like particle can give a pretty good agreement with the experimental data.

%\section{The color flow method analysis\label{colorflow}}
{\bf The Color Flow Method Analysis:} In this paper, we take the method in Ref.~\cite{Gallicchio:2010sw} to analyze the octet color of the $Z_C$. The color flow method is invented to distinguish the color of the mediate particle in quark pair production process. If we draw the transverse momentum $P_T$ of the particles of the quark pair jets in the rapidity-azimuthal angular $y-\phi$ plane, the different color of the mediate particle will lead to different shapes of the jets. For a color singlet
mediating particle, such as $\gamma/Z$, the two jet shape seems to attract to each other, while for
a color octet mediating particle, such as gluon, the two jet shape seems to repel to each other. Such a difference is caused by the different color charge flow in the Feynman diagrams. A vivid example can be found as Fig.~2 in Ref.~\cite{Gallicchio:2010sw}.

To distinguish the color flow, a measurable variate named as the Pull
is defined as~\cite{Gallicchio:2010sw}
\begin{equation}
\vec{t}=\sum\limits_{i\in {\rm jet}} \frac{P_T^i|r_i|}{P^{\rm jet}_T}\vec{r}_i,
\label{Pull}
\end{equation}
where $\vec{r}_i=(\delta y_i,\delta \phi_i)=\vec{c}_i-\vec{J}$, in which
 $\vec{J}=(y_J,\phi_J)$ is the location of the jet and $\vec{c}_i$ is
the position of a cell or particle with transverse momentum $P_T^i$.
We would like to emphasize that $\vec{t}$ is a two dimensional vector
and it can be written as $\vec{t}=(t_y, t_\phi)=(|\vec{t}|\cos \theta_t|,|\vec{t}|\sin\theta_t)$. $\theta_t$ is the polar angle determined by the two components of $\vec{t}$.

We adopt the Madgraph~\cite{Alwall:2011uj} to make the Monte-Carlo (MC) simulation. The top pair can be produced through the SM processes and the $q{\bar q}\to Z_C \to t\bar{t}$ process. We use
Pythia~\cite{Sjostrand:2006za} to simulate the top decay and the sequent hadronization to form the jets. Some special characters in the simulation are described as follows
\begin{itemize}
\item Bottom jet Pull is calculated to replace the ``top jets''. As top quark decays to bottom and the $W$ boson. Its color is fully carried by the bottom quark, and then the selection of bottom jet Pull can avoid the dilution caused by $W$ boson from the top quark.
\item When calculating the b jet Pull, a core $\vec{J}=(y_J,\phi_J)$ is needed and it is taken as the b quark from the top decay in the Monte-Carlo sample events. Such a b quark is actually invisible in the real data sample.

\item For the hadronization of the bottom quark from the top decay in the MC simulated events, the b quark firstly forms a string/cluster together with a light quark. The light quark can be from the hadron beam in the color octet mediating case or from the vacuum in the color singlet case. The string/cluster then decays into a B meson together with some light quarks which transforms to light hadrons. These light hadrons can generally be divided into two categories. The first category belongs to the beam remnant hadrons, which have large rapidity $y$ and small $P_T$. The second category is the other light hadrons which are likely to locate near to the B meson. The summed momentum of the B meson nearby hadrons and the B meson decay products are approximately equal to the momentum of the b quark from the top decay; the summed momentum of the beam line nearby hadrons is approximately equal to the initial momentum of the light quark which forms a string/cluster together with the b quark.

\item In calculating the Pull, we adopt the b quark from top decay as the core, and all the string/cluster decay products are calculated in the summation in Eq.~(\ref{Pull}). Although the first category jets actually can not be recorded in the real experiments, their contributions to the Pull are small enough to be neglected for their small $P_T$. So the involving of all the decay products from the string/cluster can still be a good approximation.

\item We find that the Pull diagrams are very sensitive to the selection of the core. For example, if we take the B meson to be the core and sum the B meson decay products in the Pull,  the color singlet or octet cases can not be distinguished. This shows that the distortion of the jet shape happens just in time of the B meson production, rather than in its decay. Once the colorless B meson is formed, the color flow is stopped. So its decay is independent of the previous color information.

\item Based on the above analysis, we can infer that the main contribution to the different Pull really comes from the light hadrons near the B meson, which maybe tagged as a part of the b jet. Experimentally, the b tagging usually requires seeing a secondary vertex from the B meson decay. Inevitably, the light quark jets will fall in the B meson cone,  are irreducible from the B meson decay products, and form a component of the b jet. Such quarks and their decay products carry the color information from the b quark, inherit from the top quark.

\end{itemize}

\begin{figure*}[htbp]
\begin{center}
\includegraphics[width=0.99\textwidth]
{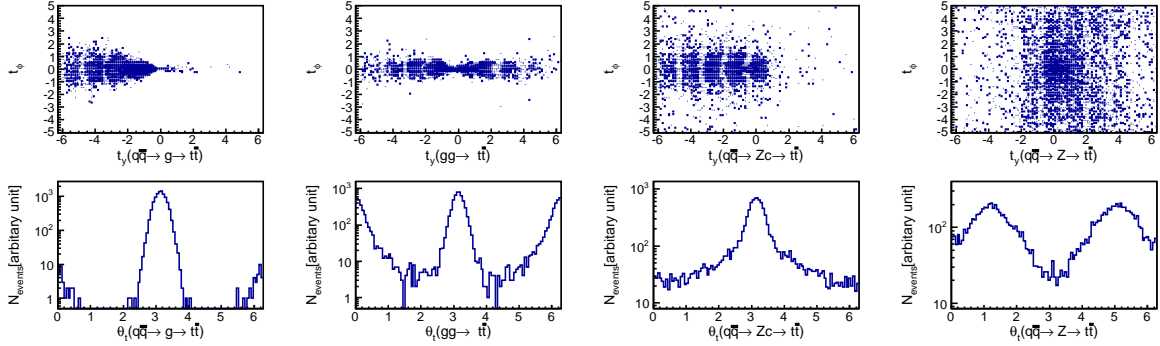}
\end{center}
\caption{\label{Tevatron_b_kernel_string}The two dimensional diagrams of the Pull $\vec{t}$ for the different color mediate particles~(upper diagrams), and the histograms of the polar angles $\theta_t$~(lower diagrams) at the Tevatron. To make the peaks just locate in the central region of the diagrams, we take the $\bar{p} p$ collide mode in the simulation. The convolution of the parton distribution function is included. }
\end{figure*}

\begin{figure}[htbp]
\begin{center}
\includegraphics[width=0.23\textwidth]
{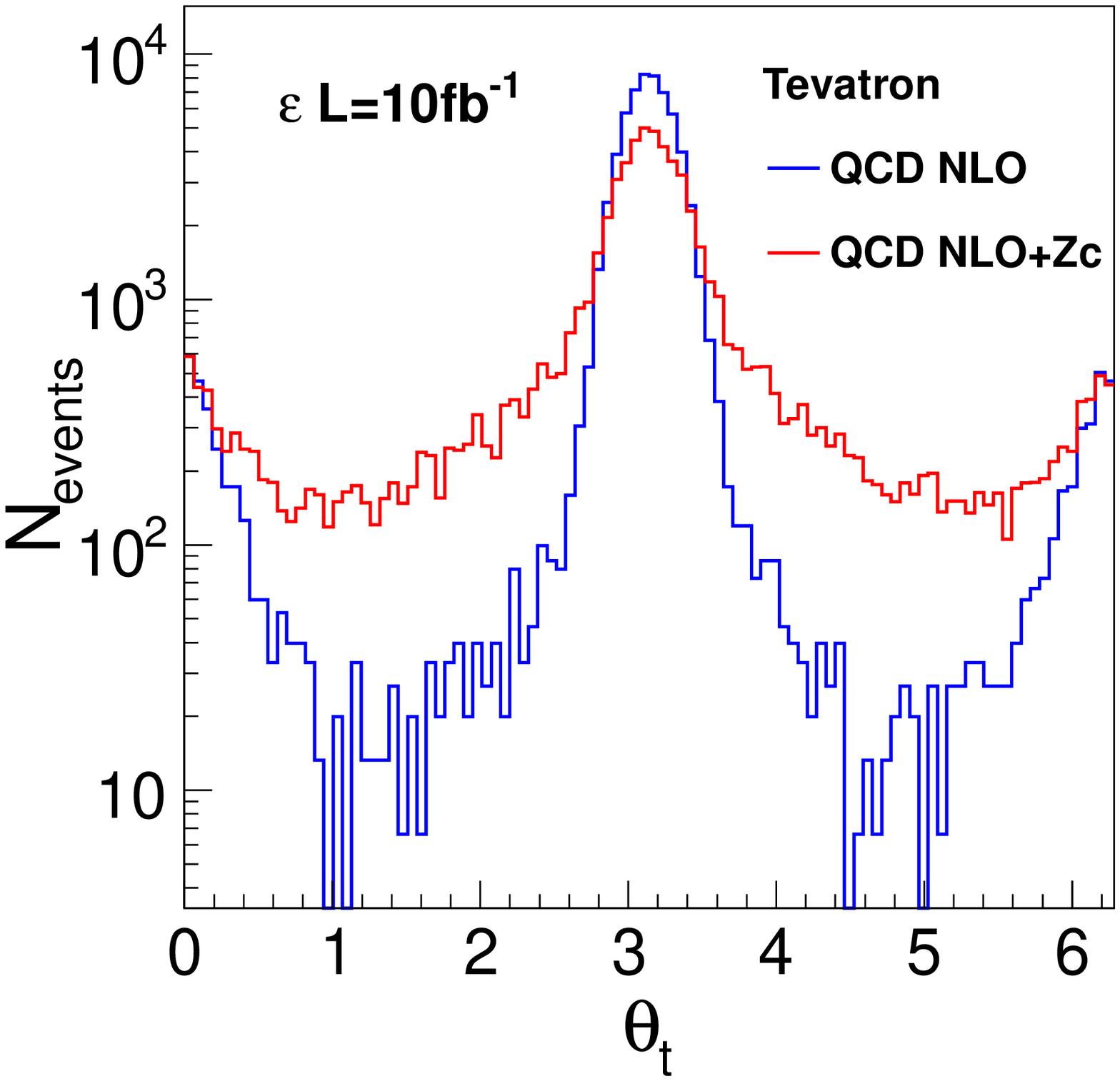}
\includegraphics[width=0.23\textwidth]
{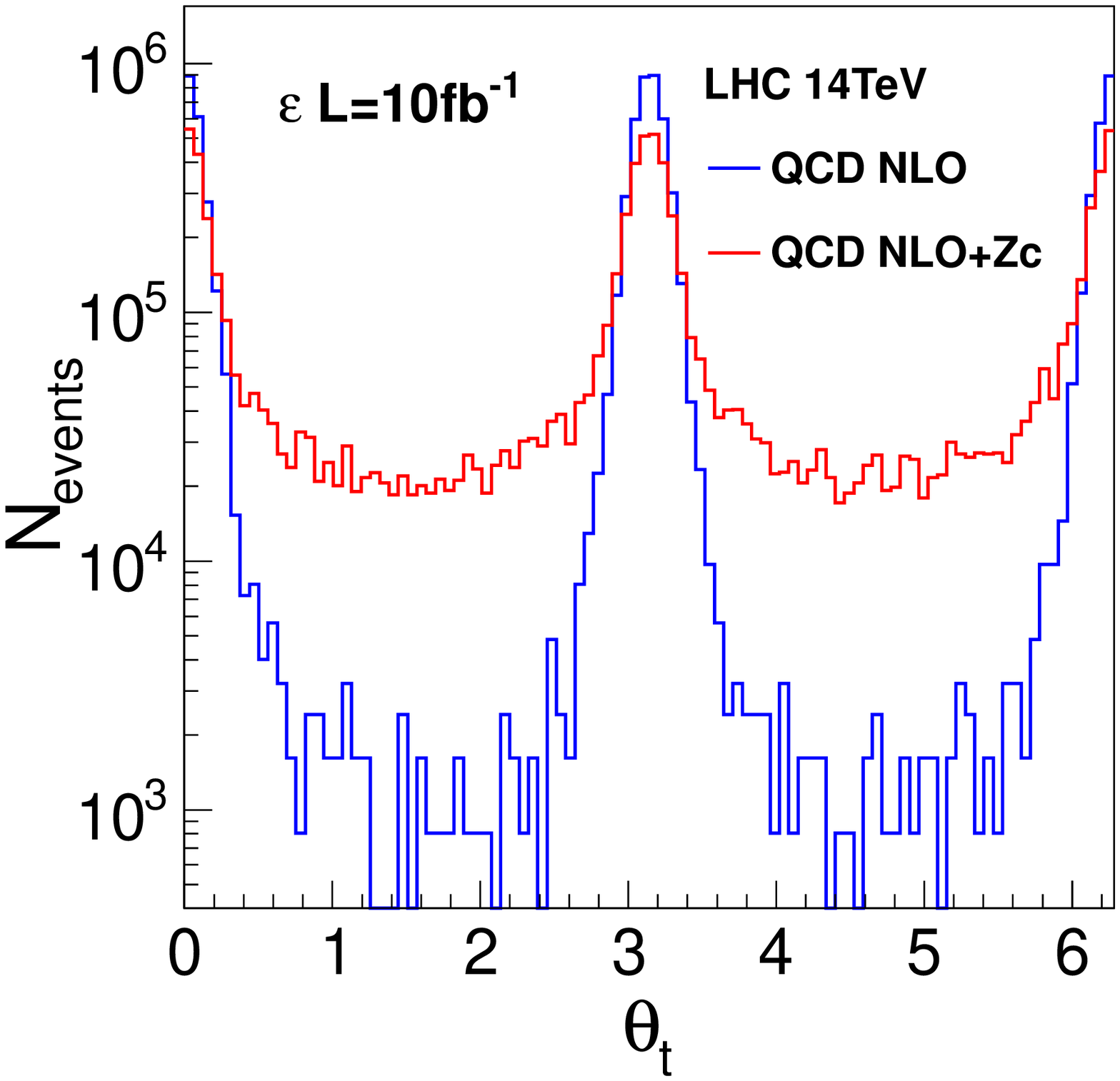}
\end{center}
\caption{\label{collider_Pull_theta_t}The distribution of polar angle $\theta_t$ of the Pull vector $\vec{t}$ at the Tevatron and LHC. $\epsilon$ is the overall event selection efficiency and ${\cal L}$ is the integrated luminosity.}
\end{figure}

Fig.~\ref{Tevatron_b_kernel_string} shows the Pull vector of the b jet and the histogram of $\theta_t$ for different processes at the Tevatron. The upper plots are the two dimensional points of the Pull, and the lower plots are the histograms of the polar angle $\theta_t$ of the Pull. It can be seen that the color octet~($g$ and $Z_C$) and singlet~(SM gauge boson $Z$) mediating particles can be distinguished clearly. For the color octet mediating particles, the b jet shape tends to repel to each other, or in other words, the color flow of the jets are connected to the remnants in the beam direction. Thus, the component of $\vec{t}$ in the rapidity $y$ direction $t_y$ is larger than the component in the azimuthal angle $\phi$ direction $t_\phi$. Consequently, the polar angle $\theta_t$ of $\vec{t}$ accumulates more in the $\pi$ or 2$\pi$ region. For the top quark produced from the $q\bar{q}$ initial state, it is more probably be attracted by the remnant beam where the $q$ comes from. That is to say, the b quark from the top decay is likely to form a string/cluster with a quark from the proton beam. This explains why $\theta_t$ accumulate more at $\theta_t=\pi$ than $\theta_t=2\pi$ in $q\bar{q}\to g/Z_C\to t\bar{t}$ process. While for the $g g$ initial state processes, the b quark from top decay has equal probability to form a string/cluster with a quark either from the proton, or from the antiproton. Thus, the peaks of the $\theta_t$ at $\pi$ or $2\pi$ for the $g g$ initial state processes are nearly the same. For the gauge boson $Z$ mediating process, the peaks of $\theta_t$ locates at $\pi/2$ and $3\pi/2$. This can be explained as follows: because the top quark is relatively heavy, the longitudinal boost of the top pair is small and then the top quark and the anti-top quark are likely to fly back to back. The b quark pair from the top decay are also nearly back to back
and their rapidities are small. The $\phi$ component of Pull $\vec{t}$ can be significantly larger than the rapidity component $t_y$. Thus, the polar angle $\theta_t$ accumulates more at $\pi/2$ and $3\pi/2$ region.

Fig.~\ref{collider_Pull_theta_t} shows the distribution histograms of the $\theta_t$ at the hadron colliders. For the Tevatron, it is a summation of the sub-processes as shown in Figure~\ref{Tevatron_b_kernel_string}. As we introduce a color octet particle, its peak also locates at the $\theta_t=\pi$ region. However, the peak becomes much wider than the SM case. By measuring the $\theta_t$ distributions at the hadron collider, the different shapes are distinguishable. This is especially useful at the LHC, in which the top asymmetry is too small to be measured precisely. Because the
LHC is a proton-proton collider without preferred initial direction, the peaks of $\theta_t$ at $\pi$ and 2$\pi$ have the same altitude.
To distinguish the two histograms in Fig.~\ref{collider_Pull_theta_t}, it is necessary to require $sig=|N_1-N_2|/(\sqrt{N_1}+\sqrt{N_2})>1$, where $N_1$ and $N_2$ are the event numbers in each bin
 for the two histograms. $sig$ gets its largest value at the $\theta_t$ peak region and the least effective luminosity can be estimated as $\epsilon{\cal L}=0.008/0.00002~\mbox{fb}^{-1}$ with the selection efficiency $\epsilon$ being about 0.001/0.00001 for the Tevatron/LHC~{\cite{CDF-dilepton-efficiency,Aad:2011yb}}.

%\section{Conclusion}
{\bf Conclusion:} In this paper, we study the color properties in the top quark pair production process. Firstly, we show that a color octet axial-vector like new particle with different coupling strengths to heavy and light quarks can provide an excellent effective explanation of
the top $A_{FB}$ anomaly discovered at the Tevatron. The color octet property of this new particle makes it very suitable to be studied by adopting the color-flow method. By defining the variate Pull vector, the jet substructure of the b quark from the top decay can be used to exploit the color of the mediating particle in the top pair production. The polar angle of the Pull vector can be measured precisely to distinguish from the SM predictions. This can be a first indirect cross check of the Tevatron top $A_{FB}$ excess at the LHC.

%\section*{ Acknowledgements}

{\bf Acknowledgements:} This research is supported in part
by the Natural Science Foundation of China
under grant numbers 11075003, 10821504, 11075194, 11135003,
and 11275246, and
by the Postdoctoral Science Foundation of China under grant
number 2012M510564 and 2012M520098.

\end{document}